%% file: main.tex
\documentclass[%
 reprint, 
 amsmath,amssymb,
 aps, prx, 
]{revtex4-2}

\usepackage{graphicx} 
\usepackage{dcolumn}
\usepackage{bm}
\usepackage{blindtext}

\usepackage{braket}
\usepackage[T1]{fontenc}
\usepackage{float}
\usepackage{xcolor}

\begin{document}

\title{Ultrafast excitonic dynamics in DNA:\\ Bridging  correlated quantum dynamics and sequence dependence}

\author{Dennis Herb}
\altaffiliation{Equal contribution}
\affiliation{Institute for Complex Quantum Systems, Ulm University, 89069 Ulm, Germany}
\affiliation{Center for Integrated Quantum Science and Technology (IQST) Ulm-Stuttgart, Germany}
\author{Mirko Rossini}
\altaffiliation{Equal contribution}
\affiliation{Institute for Complex Quantum Systems, Ulm University, 89069 Ulm, Germany}
\affiliation{Center for Integrated Quantum Science and Technology (IQST) Ulm-Stuttgart, Germany}
\author{Joachim Ankerhold}
\affiliation{Institute for Complex Quantum Systems, Ulm University, 89069 Ulm, Germany}
\affiliation{Center for Integrated Quantum Science and Technology (IQST) Ulm-Stuttgart, Germany}

\begin{abstract}

After photo-excitation of DNA, the excited electron (in the LUMO) and the remaining hole (in the HOMO) localized on the same DNA base form a bound pair, called the Frenkel exciton, due to their mutual Coulomb interaction. In this study, we demonstrate that a tight-binding (TB) approach, parametrized by ab initio data, allows to correlate relaxation properties, average charge separation, and dipole moments to a large ensemble of double-stranded DNA sequences (all 16,384 possible sequences with 14 nucleobases). This way, we are able to identify a relatively small sub-ensemble of sequences responsible for long-lived excited states, high average charge separation, and high dipole moment. Further analysis shows that these sequences are particularly T-rich. By systematically screening the impact of electron-hole interaction (Coulomb forces), we verify that these correlations are relatively robust against finite-size variations of the interaction parameter, not directly accessible experimentally. This methodology combines simulation methods from quantum physics and physical chemistry with statistical analysis known from genetics and epigenetics, thus representing a powerful bridge to combine information from both fields.

\begin{description}  
\item[Key words]
Tight-binding model, Frenkel exciton, electron-hole interaction, exciton decay, relaxation time, charge separation, dipole moment, DNA polarization, fluctuations, and epigenetics.
\end{description}

\end{abstract}

\date{\today}
\maketitle

\input{chapters/1_introduction}
\input{chapters/2_theory}

\input{chapters/3_lifetime_and_dipole}
\input{chapters/4_basic_analysis}
\input{chapters/5_lifetime_with_interaction}
\input{chapters/7_conclusion}

\begin{acknowledgments}
The authors gratefully acknowledge Marco Trenti for his assistance with numerical simulations. The authors appreciate the financial support of the Center for Integrated Quantum Science and Technology (IQST) and the BMBF through the QCOMP project in Cluster4Future QSens.
\end{acknowledgments}
\bibliography{bibliography}

\appendix
\input{chapters/appendix.tex}

\end{document}

%% file: chapters/1_introduction.tex
\section{Introduction}\label{sec:introduction}

Photoexcited energy and charge transfer in biologically relevant systems is one of the fundamental processes in nature and the basis for all life on earth. Exploring its microscopic mechanisms has been a major field of research over the past decades in biology, chemistry, and physics. One prominent example is the formation of electron-hole pairs (excitons) on complex molecular structures under external radiation. It not only serves as a model system to 
better understand molecular energy transport, e.g., in light-harvesting complexes in photosynthesis \cite{Fassioli2014, Mirkovic2017, Chenu2015, Cao2020, Mattoni2020} but has also triggered the design of nature-inspired tools such as efficient organic solar cells \cite{Ostroverkhova2016, Inganäs2018}, energy storage devices \cite{Lu2021}, and organic photodetectors and phototransistors \cite{Ostroverkhova2016}.

In particular, photoexcited charge dynamics on DNA, the central repository of genetic information in the cell, has been intensively studied because of its relevance 
to induce DNA lesions, alter DNA binding properties, and trigger carcinogenic mutations. As a result, a substantial body of literature has appeared reviewing charge transfer processes in DNA \cite{Improta2016,Crespo-Hernandez2004,Middleton2009,Kleinermanns2013,Schreier2015, Schuster2004, Endres2004} and proposing applications ranging from genetics, e.g., understanding DNA damage and repair \cite{Rajski2000,Rajski2001,Giese2006} and distinguishing pathogenic from non-pathogenic mutations \cite{Shih2011}, to nanotechnology, e.g., designing nanosensors, nanocircuits, and molecular wires \cite{Chakraboty2007,Albuquerque2014}. 

DNA is known to be photostable with rapidly decaying excitations. Improved spectroscopic methods, such as pump-probe techniques, have made it possible to experimentally study the dynamics of ultrafast excited states in DNA \cite{Crespo-Hernandez2005, Takaya2008}. These states can be generated naturally by ultraviolet (UV) radiation, for example as a result of direct exposure to sunlight. In both cases, an electron is excited from the highest occupied molecular orbital (HOMO) to the lowest unoccupied one (LUMO), leaving a hole in the HOMO. These charge carriers interact as a result of attractive Coulomb forces and, when bound to a single nitrogenous DNA base, form an electron-hole pair known as a Frenkel exciton. As a localized excitation, the Frenkel exciton plays an important role in processes such as energy transfer in DNA-based systems \cite{Wang2022,Bittner2006,Bittner2007}.     

A theoretical description of the propagation of charges through DNA is rather challenging. It is sensitive not only to extrinsic effects, such as the surrounding aqueous solution, but also to intrinsic influences, such as fluctuations in the dynamical structure \cite{Gutierrez2009, Gutierrez2010, Lambropoulos2019}, including the phosphate backbone. As a consequence, the complexity of the DNA molecule and its environment renders atomistic ab initio calculations even in equilibrium not feasible for any structures beyond 3-4 nucleotides, that is, far from capturing long-range charge transfer in double-strand aggregates. Therefore, so-called tight-binding models have been developed as powerful alternatives \cite{Cuniberti2007, Lambropoulos2019}. 

Here, we follow this strategy with the goal of combining
two methodologies, namely, mesoscopic modeling that captures the essential features of excited-state charge dynamics and a statistical analysis of sizable sets of double-strand DNA sequences. With model parameters taken from first-principle calculations \cite{Simserides2014,Hawke2010}, we achieve sufficiently accurate results with the required high numerical efficiency to screen larger ensembles.

Most of the previous studies have focused on the hole transfer along the DNA strand for various reasons: First, experiments aimed at investigating the motion of charges along DNA often make use of DNA oxidation \cite{Giese2002, Giese2006} (lack of an electron) to create an initial charge carrier; second, ab initio calculations (such as DFT) proved to be more accurate in retrieving low energy structures and properties of HOMO molecular orbitals. However, more recent literature \cite{Hawke2010} reports  a comprehensive collection of structural data such as binding energies and overlap integrals which provides the basis
to develop models including both electron and hole motion. 

In this work, we go beyond by promoting a simplistic, yet relatively accurate, unified model to analyze the dynamics of electrons and holes along DNA double strands. In contrast to similar models discussed in the literature, we refine the model by adding two crucial ingredients, namely the Coulomb interaction among both charge carriers and the finite lifetime of excitons due to local charge recombination.

Our hypothesis for this study is that there is a direct relation between effective exciton lifetime and, induced by the mobility of electron and hole along the DNA, their mean spatial separation; the latter depends on the specificities of the DNA sequences encoded in the tight-binding parameters. 
If this were true, at least for a sub-ensemble of DNA sequences of certain length, the consequences could be far-reaching: A relatively stable charge distribution resulting from charge delocalization within pockets of the DNA molecule would result in the emergence of electrical dipole forces that, in turn, may affect local conformational structures of the DNA complex and its local electrical properties. Both phenomena could in principle have an impact on the likelihood of binding
of regulatory proteins to the DNA \cite{Siebert2023}, a crucial aspect of gene regulation and transcriptional processes. 

In order to test this hypothesis, we combine efficient quantum dynamical simulations with the screening of a large set of DNA sequences. More specifically, our systematic analysis using advanced computational implementation includes up to 16,384 possible DNA double strand sequences of length seven bases such that, based on a statistical analysis, we are able to combine information from quantum physics/physical chemistry with that from genetics or epigenetics. This way, we aim to identify DNA sequences which may be of particular interest to be explored further in experimental settings of genetics/epigenetics. 
 
 Remarkably, we first find that our predictions in terms of exciton lifetime and spatial separation for the most stable species are in good agreement with data obtained in previous experiments \cite{Crespo-Hernandez2005,Genereux2010,Buchvarov2007, Takaya2008, Kadhane2008}. Our results show further that it is only a small subset of sequences that stands out as responsible for relatively long-lived excited states and high-average charge separation. The role of the electron-hole interaction is very decisive: Assuming that the Coulomb interaction between charge carriers is completely screened (no interaction) leads statistically to significantly different subsets compared to the case when it is finite. However, if it is taken as finite and is varied in a reasonable window of values \cite{Simserides2014, Lambropoulos2016a}, the interesting subset turns out to be relatively robust. This is important since the precise value of this Coulomb interaction is not known and is experimentally not directly accessible. 

Although still in the embryonic stage, we hope that the presented methodology may trigger future research by, for example, developing even more elaborate tight-binding models and optimized numerical tools based on AI \cite{Korol2019, Ullah2022} for even larger sets of DNA. 

The present manuscript is organized as follows: in Sec.~\ref{sec:model} we provide a description of the model used in our study. In Sec.~\ref{sec:lifetime_and_dipole} we systematically investigate exciton lifetime and average charge separation without e-h interaction. We introduce a Coulomb interaction between the charges in Sec.~\ref{sec:interaction} and present its effect on exciton lifetime and average charge separation in Sec.~\ref{sec:e-h_lifetime_and_dipole}. We conclude in Sec.~\ref{sec:conclusion}. 

%% file: chapters/2_theory.tex
\section{Theoretical modeling}\label{sec:model}

\begin{figure*}
\centering
\includegraphics[width=0.8\linewidth]{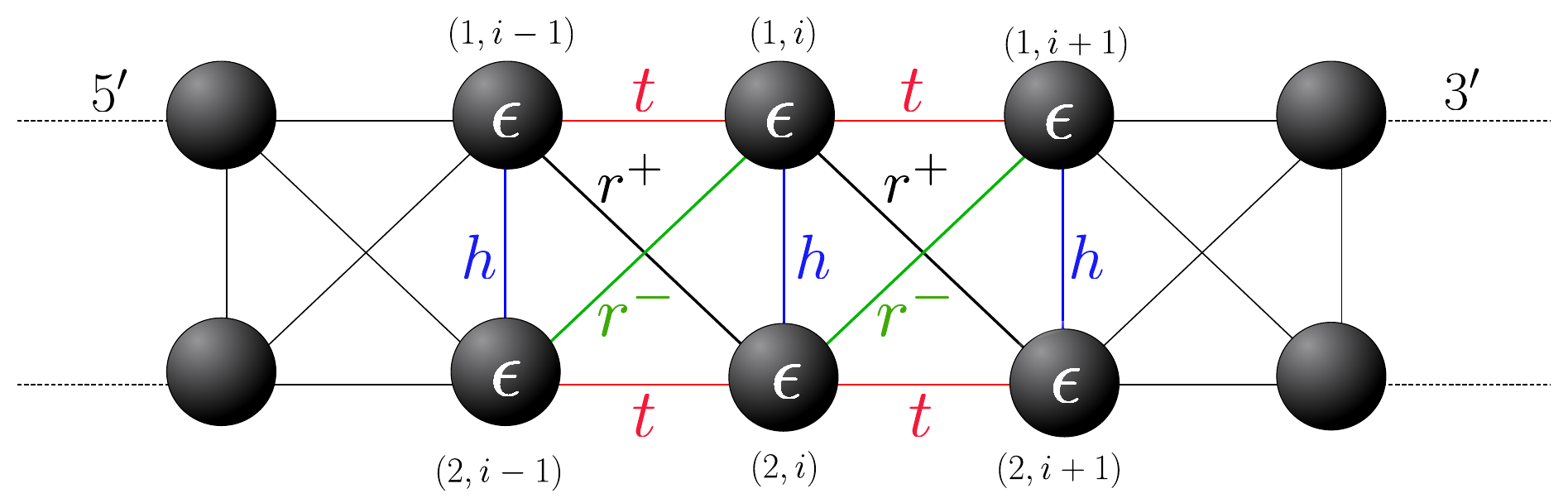}
\caption{\label{fig:1} Extended ladder model (ELM) with on-site energies and interaction parameters, including the diagonal interstrand interaction parameters $r_{\pm}$.}
\end{figure*}

We consider each of the four DNA bases (A, C, G, T) as a site in a (two-dimensional) tight-binding lattice (see Fig.~\ref{fig:1}) and numerate the sites by an index tuple $(ij)$ where $i \in \{ 1,2 \}$ refers to the upper or lower strand, and $j \in \{1,...,N\}$ denotes the base index for each DNA strand  with $N$ being the number of bases per strand. Two charges (electron and hole) are created by an initial excitation. This is reflected in the Hilbert space of the system $\mathcal{H}_{S}=\mathcal{H}_e\otimes\mathcal{H}_h$ with basis states $\ket{e_{ij}h_{kl}}$ that describe all possible electron-hole configurations on the lattice. 

The overlap of the LUMO (and HOMO) orbitals of neighboring DNA bases allows an electron (and a hole) to move and spread along the DNA molecule. The Hamiltonian for both the electron and hole dynamics can be described by two terms, $H_{\mathrm{self}}^p$ and $H_{\mathrm{trans}}^p$ (where $p \in \{ e,h \}$ describes either electron or hole), containing, respectively, the on-site and transfer energies:   

\begin{eqnarray}
    H_{\mathrm{self}}^p&=&\sum_{j=1}^N \left(\epsilon_{1j}^p\ket{p_{1j}}\bra{p_{1j}}+\epsilon_{2j}^p\ket{p_{2j}}\bra{p_{2j}}\right) \label{eq:hamilton0}\\
    H_{\mathrm{trans}}^p&=&\sum_{j=1}^{N-1}\left(t_{1j}^p\ket{p_{1j}}\bra{p_{1,j+1}}+t_{2j}^p\ket{p_{2j}}\bra{p_{2,j+1}}\right)+\mathrm{h.c.}\nonumber\\
    &+& \sum_{j=1}^N h_{j}^p\ket{p_{1j}}\bra{p_{2j}}+\mathrm{h.c.} \label{eq:hamilton1}\\
    &+& \sum_{j=1}^{N-1}\left(r_{j}^{+p}\ket{p_{1j}}\bra{p_{2,j+1}}+r_{j}^{-p}\ket{p_{1,j+1}}\bra{p_{2,j}}\right)+\mathrm{h.c.} \nonumber
\end{eqnarray}
This model is commonly known as the Extended Ladder Model (ELM) \cite{Lambropoulos2019}. We can now build the Hamiltonian for the electron and hole dynamics along the DNA double-strand as $H_e=(H_{\mathrm{self}}^e+H_{\mathrm{trans}}^e)\otimes \mathbb{I}_h$ and $H_h=\mathbb{I}_e\otimes (H_{\mathrm{self}}^h+H_{\mathrm{trans}}^h)$. 

To account for electron-hole attraction due to the Coulomb forces, we introduce two-particle interaction terms of the form: 
\begin{eqnarray}
    H_{\mathrm{int}}=\sum_{i,k=1}^2\sum_{j,l=1}^N U_{ijkl}\ket{e_{ij}h_{kl}}\bra{e_{ij}h_{kl}}.
    \label{eq:hamilton2}
\end{eqnarray}

Normally, to account for these many-body interaction terms requires investigating an exponentially large Hilbert space with respect to the dimension of the system (number of DNA bases). In this case, however, we are able to enforce the constraint that there is always either one exciton or no exciton in the system, allowing us to investigate only a quadratically large Hilbert space.

More specifically, for the interaction matrix elements we assume an algebraic spatial decay according to
$U_{ijkl}=J/(1+r/r_0)$ with $r_0=1.0\,\text{\AA}$ following \cite{Bittner2006, Bittner2007}. Since two neighboring DNA bases are separated (in the intrastrand and interstrand direction) by $D\sim3.4\,\text{\AA}$, we simplify in the tight-binding description the distance dependence to be of the form
 $r/r_0\approx (D/r_0) \cdot \left(|i-k|+|j-l|\right)$ and only include interactions between next-neighbor bases in agreement with a short-range approximation used in the literature, i.e., 
\begin{eqnarray}
    U_{ijkl}=\begin{cases}
    J&\mathrm{if}\quad (|i-k|+|j-l|)=0\\
    J/(1+D/r_0)&\mathrm{if}\quad (|i-k|+|j-l|)=1\\
    0&\mathrm{else}\end{cases}\, 
    \label{eq:coulomb}
\end{eqnarray}
The first case describes the two charges being located at the same base, while the second holds when electron and hole are sitting on neighboring bases. In this way, we arrive at the interacting TB Hamiltonian for the ELM: $H=H_e+H_h+H_{\text{int}}$. The parameters for this model are taken from ab initio simulations in Ref.~\cite{Hawke2010} and are given in the Supplementary material \cite{Supplementary}. 

Typical energies that are needed to create an excitation on the DNA are given by the energy difference between HOMO and LUMO and range between $3.5\,\mathrm{eV}$ and $4.5\,\mathrm{eV}$ \cite{Hawke2010}. The typical temperature $T\approx310\,\mathrm{K}$ for DNA under ambient conditions instead corresponds to energies in the range of the meV, much too small to spontaneously induce excitons, thus reflecting the stability of native DNA.
Hence, external stimuli give rise to excitonic phenomena. In this sense, in a minimal setting, intrinsic loss mechanisms, i.e.\ exciton recombination, are the only relevant impact of residual degrees of freedom to which energy from the charge sector can be dissipated. In the spirit of quantum optical modeling, these can be described by operators of the form
\begin{equation}
    A_{ij}=\sqrt{\gamma_{ij}}\ket{0}\bra{e_{ij}h_{ij}}\, 
    \label{eq:recombination}
\end{equation}
with $\ket{e_{ij}h_{ij}}$ denoting an electron $e$ (hole $h$) at site $i$ on strand $j$ and local (monomer) transition rates $\gamma_{ij}$. 
They act as dissipators with the tendency to destroy quantum coherences in a Lindblad-type equation for the reduced density operator of the exciton, namely, 
\begin{equation*}
    \frac{\mathrm{d}}{\mathrm{d}t}\rho(t)=-\frac{\mathrm{i}}{\hbar}[H,\rho]+\sum_{i, j} A_{ij}\rho A_{ij}^{\dagger}-\frac{1}{2}\left(A_{ij}^{\dagger}A_{ij}\rho+\rho A_{ij}^{\dagger}A_{ij} \right).
\end{equation*}
Here, the total Hamiltonian enters the deterministic part while the second part describes local exciton recombination to the monomer groundstate, see Fig.~\ref{fig:2} (for further details see below). 

The above time evolution equation is now represented in the above site basis. Time-dependent expectation values are obtained from the respective density operator $\rho(t)$ at time $t$. For example, the population in the electron (hole) sector only follows from 
\begin{equation*}
    P_{ij}^p(t)=\bra{p_{ij}}\rho_p(t)\ket{p_{ij}}\   
\end{equation*}
with the partial trace $\rho_{p} = $ Tr$_{p'}[\rho], p\neq p'\in \{e, h\}$. 
The exciton population at the same site is given by 
\begin{equation*}
P_{ij}^{ex}(t)=\bra{e_{ij}h_{ij}}\rho(t)\ket{e_{ij}h_{ij}}
\end{equation*}
To incorporate the possibility for e-h recombination, the underlying Hilbert space is expanded by adding the state $\ket{0}$ to the set of above basis states. It represents the ground state of the DNA sequence, as illustrated in Fig.~\ref{fig:2}, with a population given by $P_0(t)=\bra{0}\rho(t)\ket{0}$.

Based on this modelling,  the effective exciton lifetime as well as the charge separation on the lattice is determined. For the latter, we introduce the charge separation operator $\hat{d}$
\begin{equation*}
    \bra{e_{ij}h_{kl}} \hat{d} \ket{e_{ij}h_{kl}} = D\cdot(|i-k|+|j-l|). 
\end{equation*}
whose time-average is an estimate for the amount of charge separation during the dynamics, i.e. 
\begin{equation*}
    \bar{d}=\frac{1}{T}\int_0^{T} \ {\rm Tr}\{\rho(t)\, \hat{d}\}. 
\end{equation*}
When the electron and hole are temporarily separated in space, they form an electric dipole with an effective dipole moment, which is directly determined by the average charge separation. 

\begin{figure}[h]
\centering
\includegraphics[width=1.0\linewidth]{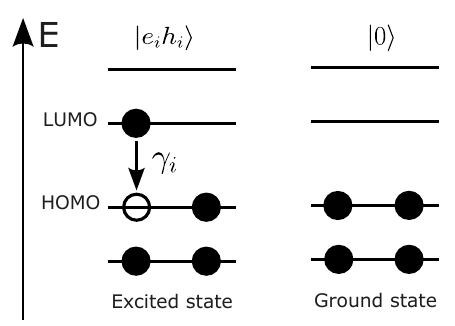}
\caption{\label{fig:2} HOMO and LUMO for excited state and ground state of the DNA molecule. Whenever electron and hole are located at the same base they recombine with a rate $\gamma_i$ such that the DNA relaxes to its ground state.  }
\end{figure}

%% file: chapters/3_lifetime_and_dipole.tex
\section{Results}
\subsection{Excited-state lifetime and average dipole moment}\label{sec:lifetime_and_dipole}

\begin{figure*}
\centering
\includegraphics[width=1.0\linewidth]{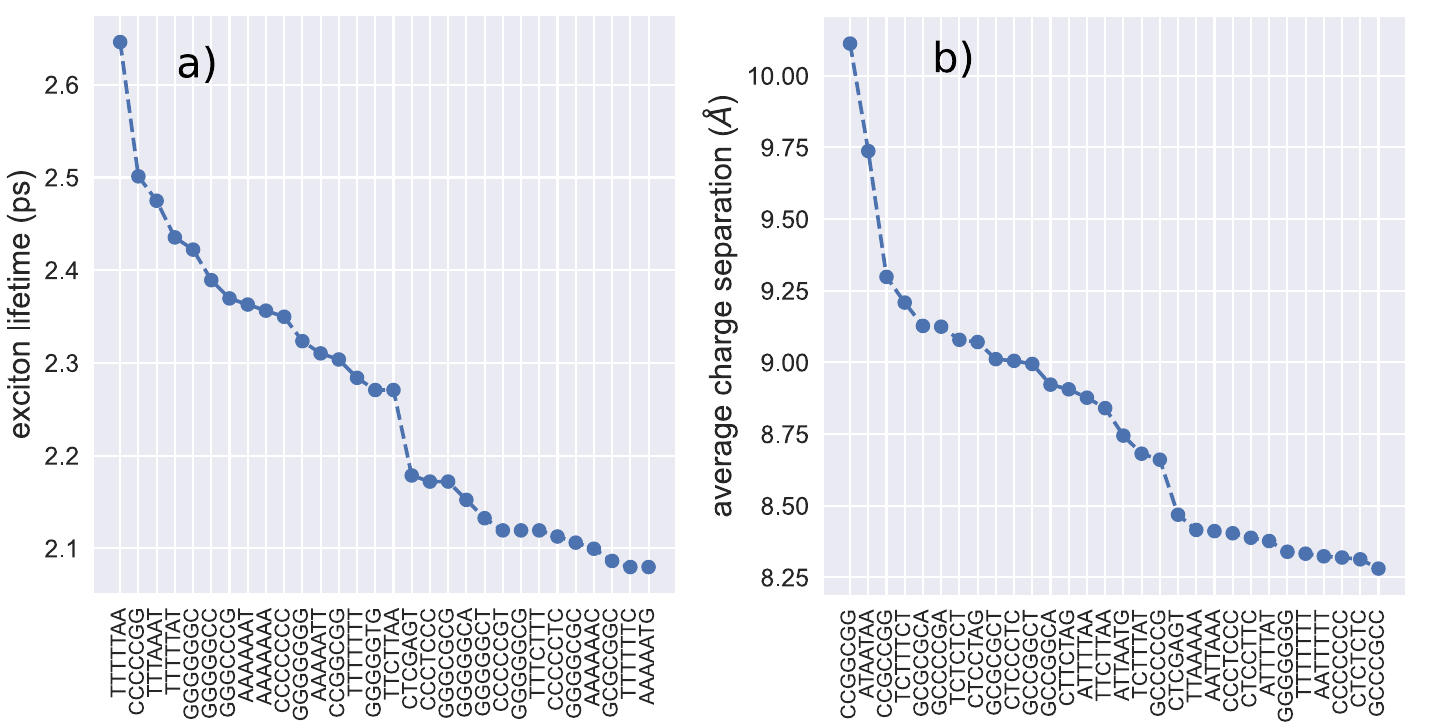}
\caption{\label{fig:3} Top 30 DNA sequences with the longest exciton lifetime (subset $S_{\rm time}$) (a) and the highest average charge separation (subset $S_{\rm space}$) (b) from the analysis of all possible double-strand sequences with seven DNA base pairs for each strand $S_7$. We assume a uniform relaxation rate $\gamma=2\,\mathrm{meV}/\hbar$ and perfect charge screening (no e-h interaction $J=0$). The letters denote the DNA bases on the upper strand in the 5'-3' direction.}
\end{figure*}

In this section, we explore the exciton dynamics with particular focus on lifetime and spatial separation, which in turn allows to reveal their correlations with respect to DNA sequences.
 Our analysis of exciton lifetimes as such appears to be of relevance in the context of pump-probe experiments presented in the literature \cite{Crespo-Hernandez2004, Crespo-Hernandez2005, Trifonov2005}, thus addressing the yet not fully understood topic of relatively long-lasting charge-hole excitations (in the range of ps) after photo-exciting specific DNA strands. 
The second observable, related to exciton mobility, provides information about the charge distribution along  DNA strands and sheds light on the question over which length scales (number of bases) quantum coherences may survive, a topic that has been discussed but not yet conclusively answered in previous work \cite{Genereux2010,Boon2002}. In fact,  possible implications are far-reaching since long-range spatial separation of electron-hole pairs may give rise to relatively strong electrical dipole forces, which in turn may open pathways for conformational changes or modify the binding affinity for CpG methylation or other enzymatic reactions.

More specifically, we conducted a systematic screening of lifetime and charge separation for all double-strand DNA sequences ranging between $N = 3$ and $N = 7$ bases in each strand (i.e. total number of bases between 6 and 14), thus covering up to $S_7\equiv 4^7=16,384$ DNA sequences. Each sequence has its individual footprint in form of a specific set of parameters for the ELM, see Eqs.~\eqref{eq:hamilton1}, \eqref{eq:hamilton2}, obtained from atomistic calculations \cite{Hawke2010, Simserides2014}. Monomer exciton recombination rates in Eq.~\eqref{eq:recombination} are not known theoretically and have thus been inferred from pump-probe experiments in \cite{Crespo-Hernandez2005}. There, values have been reported in the range of hundreds of femtoseconds for the recombination of an exciton confined within a single DNA base. We have then chosen to assume a constant rate for all modeled bases corresponding to an average lifetime of 300 fs. Variations on these time scales do not significantly affect the quantitative results for lifetime and charge separation, and thus have only a marginal impact on relative dependencies between different DNA sequences (see Supplementary material \cite{Supplementary}). 

We observed that exciton lifetimes vary substantially within the total set $S_N$ of DNA sequences, with several sequences showing lifetimes on the order of picoseconds (e.g., $>2.5\,\text{ps}$ for TTTTTAA) versus many other sequences showing lifetimes comparable to the monomer lifetime of approximately $300\,\text{fs}$ (see Supplementary material \cite{Supplementary}). The range between the longest and shortest lifetimes spanned nearly one order of magnitude. Data of the top 30 sequences (ordered accordingly in a subset $S_{\rm time}\subset S_7$) are shown in Fig.~\ref{fig:3}(a) with typical lifetimes of the order of 2 ps and a variation of approximately 25\%.

The charge separation analysis revealed a wide range of results among different sequences, with some exhibiting negligible separation, while others displaying charge separations exceeding $10\,\text{\AA}$, equivalent to an average separation of approximately three base pairs.
Following a similar analysis to the one above, we identified an ordered set $S_{\rm space}\subset S_7$ of 30 sequences with variations of approximately 20\%, see Fig.~\ref{fig:3}(b). Notably, charge separations within $D_{\rm large}$ qualitatively match previous findings in \cite{Buchvarov2007, Takaya2008, Kadhane2008, Genereux2010}, where a mixed diffusion mechanism for charge transfer along DNA strands is proposed in which quantum delocalization plays a relevant role in patches of up to $3 - 4$ base pairs, i.e.\ approximately $\sim 10\,-\, 13\, \text{\AA}$. 

Now, based on this analysis we turn to consider correlations between lifetime and charge separation. Indeed, within the total set $S_7$, we find robust correlations  with correlation factor exceeding $\chi\gtrsim 0.95$ (see Supplementary Material \cite{Supplementary}). However, when considering only the respective top subsets  $S_{\rm time}$ and $S_{\rm space}$ a remarkably weaker correlation appears: less than $\chi\approx 0.4$. For these specific sequences,  a strong correspondence between lifetime and dipole moment is not clearly recognizable. The first finding is intuitive: 
 Excitonic recombination is substantially inhibited when electron and hole sit on different sites (bases) as they are then efficiently screened by the surrounding background.
 This implies that larger charge mobility along the chain leads in general to a longer excitonic lifetimes. The second finding suggests though that details are more subtle and, among the top sequences, additional mechanisms besides the mere particle separation play a crucial role such as onsite energies.

\begin{figure*}
\centering
\includegraphics[width=1.0\linewidth]{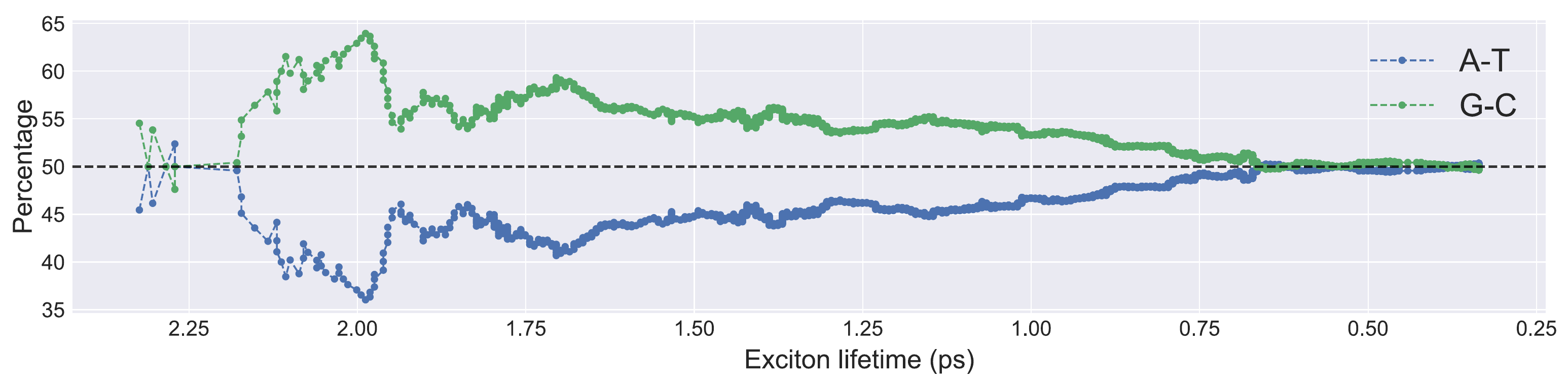}
\caption{\label{fig:4} Relative distribution of base-pairs (A-T pair and G-C pair) in sequences of seven base pairs over the expected exciton lifetime. We obtain the distribution at time $t$ by counting the base pairs in all sequences with a lifetime greater than $t$ and dividing by the number of base pairs. Since it is necessary to average over several sequences before the results become relevant, we have omitted the first 10 sequences with the largest exciton lifetimes.}
\end{figure*}

A general trend can be found when relative distributions of A-T versus G-C base pairs in $S_7$ are studied (see Fig.~\ref{fig:4}). 
Sequences containing regions rich in G-C base pairs are more likely to feature long-lived excitations than sequences rich in A-T base pairs.
In all these data, increasing (decreasing) monomer relaxation rates within a window around $300\,\mathrm{fs}$ leads to a mere shift of absolute values towards smaller (larger) exciton relaxation times but overall preserving relative relations.

%% file: chapters/4_basic_analysis.tex
\subsection{Electron-hole interaction and charge distribution} \label{sec:interaction}

While single charge transfer along DNA has been largely investigated \cite{Lambropoulos2016a, Lambropoulos2016b, Simserides2014}, the role of charge-charge interactions has received much less attention.  In the context of the current work, this is particularly true for electron-hole correlations due to attractive Coulomb forces. In fact, to include this interaction may be decisive for the dynamics of excitons as it implies to account for two competing processes: Electron-hole interaction with the tendency to localize charge pairs on individual sites (nucleotides) and the mobility of both charge carriers in LUMO and HOMO with the tendency for delocalization. Details thus depend on energy profiles, strength of hybridization, and screening properties (range of Coulomb interaction) of a specific DNA sequence. 

Here, we start this analysis considering the impact of electron-hole correlations within the so-called wire model \cite{Cuniberti2007}. The latter is a simplified version of the ELM in which the bases forming a base pair are treated as one macro-molecule of two adjacent molecules with electronic overlap such that HOMO and LUMO are localized only on one of the bases of a pair \cite{Hawke2010}. 
This model has been previously studied in the literature and thus provides benchmark data that we used to validate our modeling. It further allows to achieve a transparent dynamical understanding of electron-hole interaction which will serve as the basis to consider it in the full ELM in the next section.

Coulomb forces are usually considered as well-established and well-known interactions, whose intensity can be easily evaluated given specific boundary conditions. However, the situation in macro-molecular aggregates such as DNA is more complex, though, because the strength of this interaction can vary due to several factors inherent to surrounding degrees of freedom. An aqueous environment acts as a polarizable medium (water has a relatively high dielectric constant of about 80 at physiological temperatures), thus influencing and screening the Coulombic forces in various ways. Additionally, factors such as local hydration and the presence of counter-ions can contribute to variations in the effective dielectric constant of the surroundings. Structural deformations, which encompass mechanical stress, temperature changes, and interactions with surrounding biomolecules or macro-molecules, as well as conformational changes such as bending or twisting of the DNA helix, can further impact the spatial arrangement of the charge carriers and modify the strength and range of the Coulombic forces. Consequently, the strength of the Coulomb interaction within DNA can exhibit considerable variability. Due to these reasons, in our study, we investigate a comprehensive range of interaction strengths, ranging from considerably weaker to stronger with respect to the DNA hopping energies.

\begin{figure*}
\centering
\includegraphics[width=0.9\linewidth]{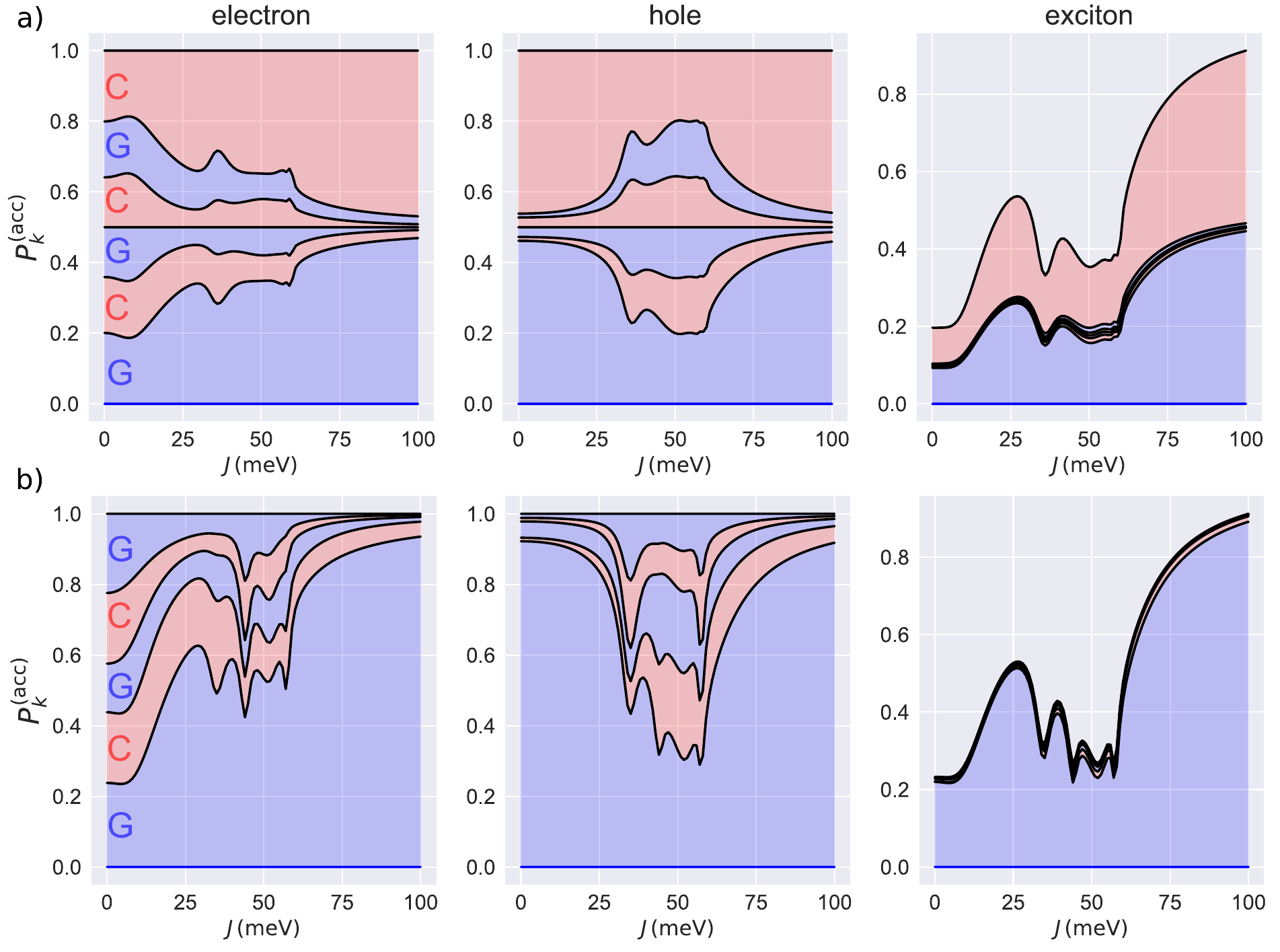}
\caption{\label{fig:5}  Accumulated average populations $P_k^{\text{acc}}=\sum_{i=0}^k\langle P_i\rangle$ of (a) GCGCGC and (b) GCGCG with initial excitation on the left base pair according to the wire model description. The hopping parameters are as follows: GC $-10\,\mathrm{meV}$, CG $-8\,\mathrm{meV}$ for the electron and GC $-10\,\mathrm{meV}$, CG $50\,\mathrm{meV}$ for the hole. }
\end{figure*}

More specifically, we investigate the effect of e-h interaction on time-averaged populations for GC periodic sequences, modeled as one-dimensional chains whose base units model the G-C DNA nucleobase pair. A similar analysis is shown in the Supplementary Material \cite{Supplementary} (Fig.~3), where  we report results for a uniform G sequence. Since the base units of the chain are the same (G-C nucleobase pairs), we note that the difference between the alternating GC sequence and the uniform G sequence lies in the staggered (or uniform) site-site overlap parameters along the chain. 
Accordingly, within this wire model, this sequence becomes degenerate with respect to on-site energies but not with respect to coupling energies as the 5'-3' DNA directionality implies that the G-C molecular overlap is different from C-G one. When we set the e-h interaction strength to zero, i.e.~ $J=0$ in Eq.~(\ref{eq:coulomb}), our results match the findings of the above-mentioned works \cite{Lambropoulos2016a, Lambropoulos2016b, Simserides2014}: The electron spreads over all DNA bases, while the hole is either distributed between the left and right edges (consider $J=0$ for even-length chains, Fig.~\ref{fig:5}(a)) or localized on the left edge ($J=0$ for odd-length chains, Fig.~\ref{fig:5}(b)). This different behavior can be easily explained by considering the physics of one-dimensional topological systems, such as the Su-Schrieffer-Heeger (SSH) chain \cite{Asbóth2016}. By inspecting the interaction energies of electrons and holes in Tab.~VII in the Supplementary material \cite{Supplementary}, one notices that in the electronic sector one has $|t_{GC}| > |t_{CG}|$, while the opposite applies to holes. In SSH chains this in turn induces two different regimes, namely a \textit{trivial} and \textit{topological} one, characterized by exactly the two dynamical scenarios seen for electrons and holes here.

Now, when switching on the e-h interaction $J\neq 0$, we observe the tendency to wash out these differences, and two different regimes can be distinguished: In the intermediate regime ($J$ is of the same order as the site-site overlap parameters), the electron population at the edges increases slightly while the hole spreads across the molecule and populates the bases in the bulk, very similar to the behavior of the electron in the uncorrelated regime. While the excitonic population (population of electron and hole sitting at the same base) thus increases, the spreading of the hole along the chain is suppressed. With increasing interaction strength, the role of electron and hole switch, with the hole pulling the electron towards the edges. In this strongly interacting regime, both charges display a similar localized distribution, leading to the observed increase in the excitonic population. 

Interestingly, we can identify an even-odd effect
by comparing results shown in Fig.~\ref{fig:5}(a) and Fig.~\ref{fig:5}(b). Sequences with an even or odd number of sites (base pairs) show a rather different behavior in the hole sector. In "even-length" sequences, the hole tends to be distributed at the edges of the strand, not populating the bulk, in both cases when it is not interacting with the electron and when it does so strongly. The same is true for the excitonic distribution in the strong interaction regime. In contrast, in "odd-length" sequences, both charges tend to localize only at the very first site of the chain, where the exciton has been created by an external stimulus. Different studies on the delocalization of quantum particles in "SSH-like" topological systems confirm this even-odd duality that we observe in this context \cite{Asbóth2016, Cramer2004}. These findings support the fact that the mobility of charge carriers and of more elaborate structures like Frenkel excitons, is way more likely to be influenced by long-range transport when placed in a wire-sequence with an even number of sites rather than an odd number, in agreement with \cite{Lambropoulos2019}. Nonetheless, both cases show very similar values for the total excitonic population along the strand.

%% file: chapters/5_lifetime_with_interaction.tex
\subsection{Electron-hole interaction and exciton lifetime}\label{sec:e-h_lifetime_and_dipole}

We now come back to the statistical analysis presented in Sec.~\ref{sec:lifetime_and_dipole} and extend it by including electron-hole interaction in the ELM. Inspired by the discussion in the previous Section, as illustrative scenarios two specific values for the interaction parameter, namely, $J=50\,\mathrm{meV}$ and $J=100\,\mathrm{meV}$ are chosen. 

\begin{figure*}
\centering
\includegraphics[width=1\linewidth]{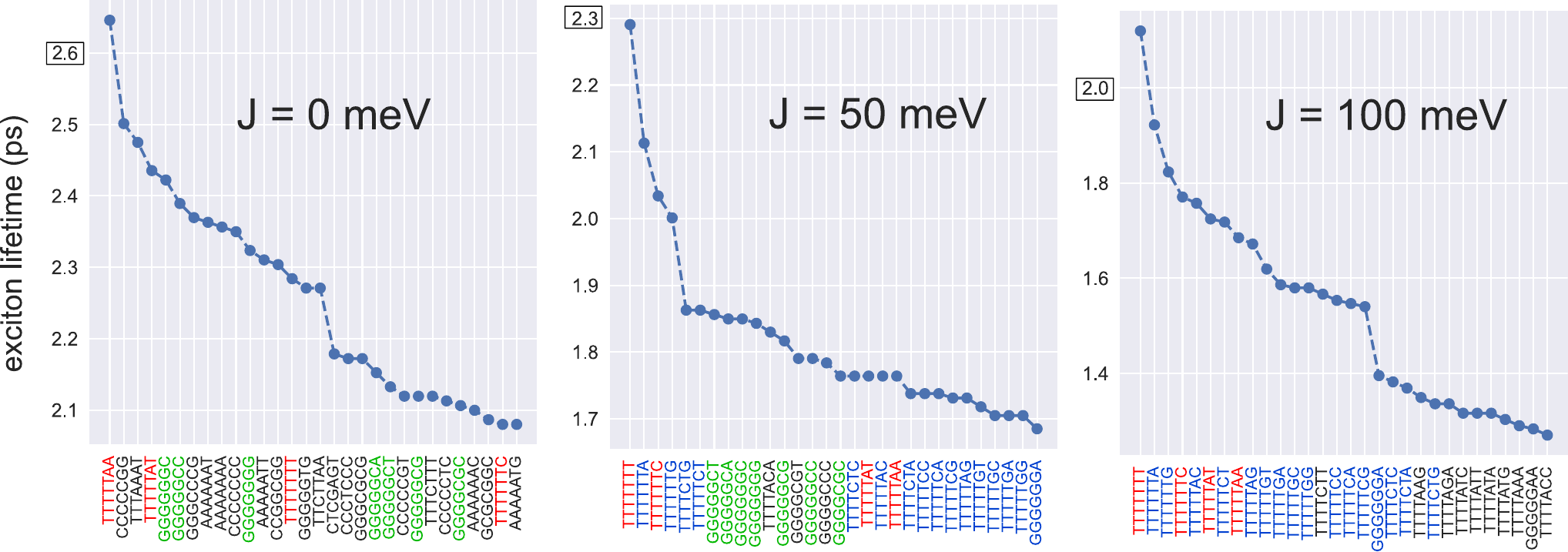}
\caption{\label{fig:6}  Expected values of excitonic lifetimes for different values of the e-h interaction strength: $0\,\mathrm{meV}$, $50\,\mathrm{meV}$ and $100\,\mathrm{meV}$. The colors of the sequences indicate the copresence of the sequence in different plots. Red: sequence present in all three plots; Green: sequence present in the first two plots; Blue: sequence present in the second and third plots.}
\end{figure*}

Fig.~\ref{fig:6} shows the top 30 sequences with the longest excitonic lifetimes $S_{\rm time}^{J\neq 0}$. One immediately sees the tendency that lifetimes are slightly reduced for all sequences compared to the situation with $J=0$ (see also the Supplementary material \cite{Supplementary}). This is easily understood by the fact that an attractive electron-hole interaction promotes local exciton recombination. 
However, quite unexpected is the finding that the sets $S_{\rm time}^{J\neq 0}$ of DNA sequences for finite $J$ are strongly correlated while correlations to the set for $J=0$ ($S_{\rm time}$) are considerably weaker. This in turn implies that the mere insertion of a finite charge-charge correlation is relevant (within a reasonable window) but not the precise value of the interaction parameter. Since the latter is not directly accessible experimentally, estimated values may already provide sufficiently reliable predictions for sequence dependences.  

\begin{figure*}
\centering
\includegraphics[width=1\linewidth]{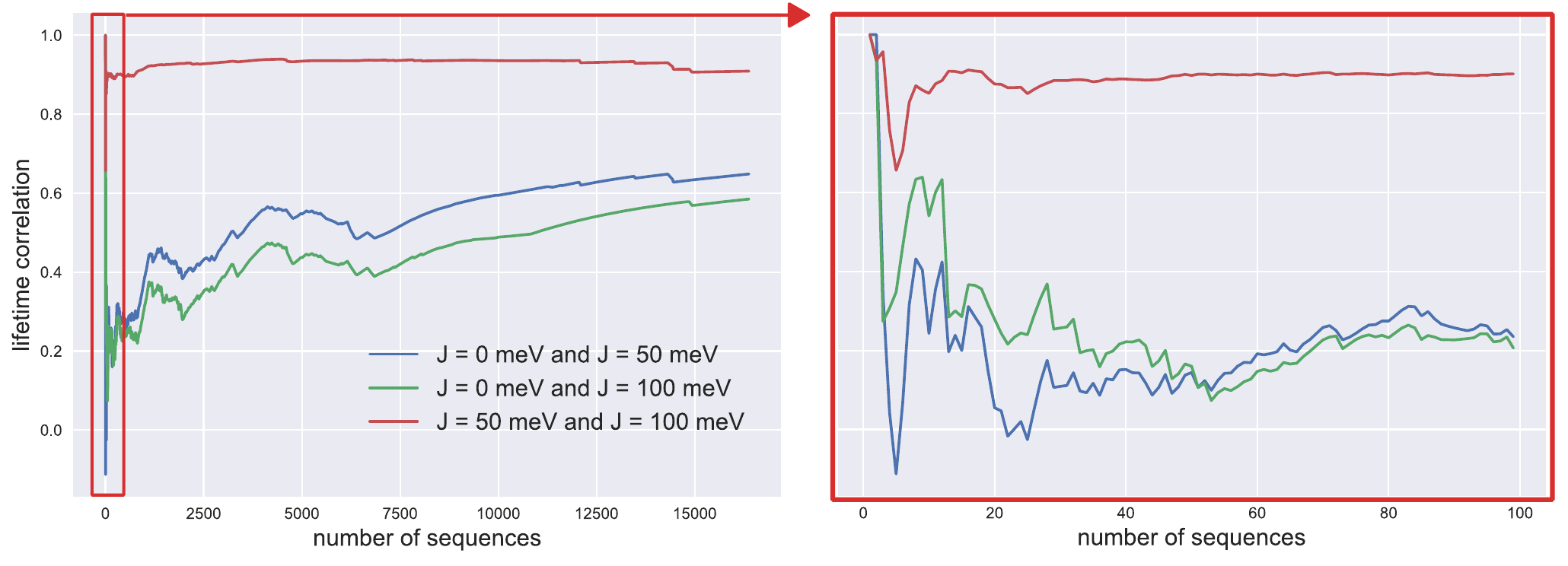}
\caption{\label{fig:7}  Correlation plots between lists of lifetimes of DNA sequences, for different values of the e-h interaction strength (see legend). The right plot is a zoom into the first 100 sequences of the left plot. The correlation between different non-zero e-h interactions is always higher than the correlations with zero e-h interactions.}
\end{figure*}

\begin{figure*}
\centering
\includegraphics[width=1\linewidth]{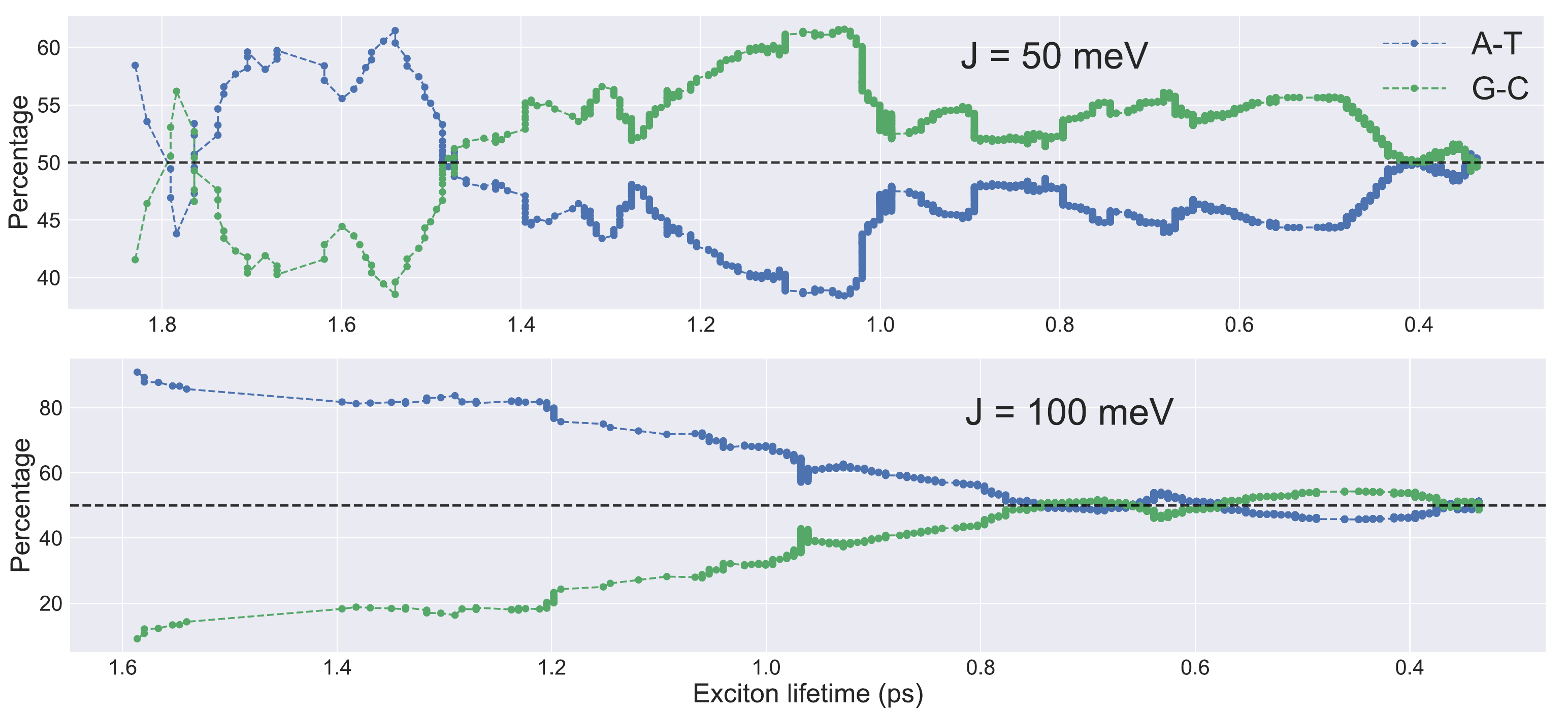}
\caption{\label{fig:8} Relative distribution of base pairs (A-T pair and G-C pair) in sequences of seven bases per strand during the expected exciton lifetime, for the e-h interaction = 50 (top) or 100 (bottom) meV. We obtain the distribution at time $t$ by counting the base pairs in all sequences with a lifetime greater than $t$ and dividing by the total number of base pairs. Since it is necessary to average over several sequences before the results become relevant, we have omitted the first 10 sequences with the largest exciton lifetimes. }
\end{figure*}

Details of this analysis are shown in Fig.~\ref{fig:7}. The left panel shows correlations for the entire list $S_7$ of DNA sequences discussed already in Sec.~\ref{sec:lifetime_and_dipole}. The correlations between the non-interacting and either of the two interacting lists (blue and green) are weak to absent within the subsets $S_{\rm time}$ resp.~ $S_{\rm time}^{J\neq 0}$ of the longest lifetimes, but moderate between the subsets of shorter lifetimes. As mentioned above, most strikingly, the correlations between lists with finite $J$ are, instead, permanently very strong, independent of the value of $J$. Even in the right panel of Fig.~\ref{fig:7}, where data for the 100 DNA sequences with the longest lifetimes are shown, correlations lie constantly above $\chi \sim 0.9$, except for the top sequences where still $\chi > 0.7$.
Very similar results are obtained by performing the same analysis for the mean charge distribution as reported in Fig.~S4 of the Supplementary Material \cite{Supplementary}.

We thus conclude that, with respect to the observables under analysis, considering the effects of e-h interactions is crucial for a realistic description of the dynamics of charges and excitons along DNA strands. The mere introduction of electron-hole interactions into the model seems to be much more relevant than the actual tuning of interaction strength.

In Fig.~\ref{fig:8} we report the same analysis as in Fig.~\ref{fig:4} for the {\rm relative} distribution of base pairs over the excitonic lifetime, for the two scenarios with finite e-h interaction. We find that the effect of the e-h interaction translates into a much greater influence of the TA base pair to ensure a higher lifetime of an exciton within a given sequence. This reversal of tendency also appears in Fig.~\ref{fig:6}, when considering the sequences that newly appear in $S_{\rm time}^{J\neq 0}$  as compared to the set with $J=0$. This suggests that T-rich sequences may more strongly support long-lived dipole structures and may thus be more susceptible to reveal properties that are of potential relevance for epigenetic processes due to photo-excitation (e.g. UV radiation in skin cells).

%% file: chapters/7_conclusion.tex
\section{Discussion}\label{sec:conclusion}

In this work we established a methodological bridge between quantum dynamical simulations of electron-hole charge transfer along double-stranded DNA and the statistical analysis of a large ensemble of DNA sequences. Although single charge transfer has been studied quite frequently, correlated electron-hole dynamics including recombination has rarely been investigated \cite{Bittner2006, Bittner2007, Tornow2010, Conwell2008}. By using a tight-binding model with parameters taken from ab initio calculations, we are able to obtain results for all possible permutations of DNA segments of up to 7 bases in each strand, far beyond what is feasible in atomistic calculations. The efficiency of this method allows us to correlate specific physical parameters to the entire ensemble of corresponding DNA sequences, thus to establishing relative dependences between them. This way, predictions can be made for subsets that support relatively long-lived, delocalized charge-hole pairs which, in turn, imply relatively strong electrical dipole moments over distances of 3-4 base pairs. 
Notably, these sequences display a particular enrichment in thymine (T) bases. The resulting charge imbalance within the DNA molecule induces molecular polarization, which has the potential to influence various biological processes, such as sequence- and conformation-dependent protein binding and transcriptional regulation. 

Our findings are quantitatively in agreement with experimental data in terms of exciton lifetimes (about 2-3 ps) \cite{Crespo-Hernandez2004, Crespo-Hernandez2005, Trifonov2005} and coherence lengths (about 10 \AA)  \cite{Genereux2010,Boon2002}. However, the tight-binding model presented is still too simplistic to allow quantitative predictions for {\em individual} DNA sequences, e.g., including a realistic description of the DNA backbone, but it does allow the model to characterize {\em relative} correlations within an ensemble of sequences.
In particular, it turns out that most subsets are relatively robust against variations of the charge-hole interaction parameter, as long as it remains finite, within a broad window of values. 

Finally, we want to discuss some implications that can be cast into the following questions: 
Might some DNA-binding proteins use DNA-mediated (correlated) charge transfer (CT) for long-range signaling or activation? CT chemistry provides an approach to detect base stacking perturbations and lesions: Might nature take advantage of this chemistry? Telomeric DNA (the region of DNA at the ends of all linear chromosomes) and regions flanking protein-coding exons are guanine-rich: Could the higher degree of degeneracy (due to repeated Gs) in these areas facilitate the tunneling of excess charges to these non-coding regions, thus avoiding dangerous complications (as speculated in \cite{Delaney2003})? 
We believe that the methodology presented in this work and its extension to larger system sizes resp. larger ensembles of DNA sequences may open new avenues to address these questions from a perspective which combines quantum dynamics with the large scale screening of DNA base compositions.
Thus, this research may complement existing techniques from physical chemistry, bio-chemistry and genetics to achieve a broader understanding of  properties of DNA with potential applications in fields such as nanotechnology \cite{Albuquerque2014}, epigenetics \cite{Siebert2023} and DNA-related studies including DNA damage, repair and mutagenesis.

%% file: chapters/appendix.tex
\section{Tight-binding parameters}

Tight-binding parameters for electrons and holes calculated using ab initio methods have been proposed in several works in the literature \cite{Mehrez2005,Hawke2010,Simserides2014}. We have decided to use the set proposed by Hawke et al. \cite{Hawke2010} for the description at the single-base level (wire model) and the set proposed by Simserides \cite{Simserides2014} for the description at the base pair level (extended ladder model). 

The LUMO and HOMO energies of the stacked DNA bases characterize the on-site energies for electrons and holes. These energies differ from the on-site energies of the isolated bases as stacked bases get slightly deformed due to base-base interactions. The energy needed to create an exciton (that means providing $\Delta E$ to excite an electron from the HOMO to the LUMO) is given by the energy difference between HOMO and LUMO. 

\subsection{Extended ladder model}

\begin{table}[H]
\caption{\label{tab:1} HOMO and LUMO energies of paired DNA bases. The HOMO energies are taken as on-site energies $\epsilon_h$ of the holes and the LUMO energies as on-site energies $\epsilon_e$ of the electrons. The energy difference $\Delta E$ between HOMO and LUMO equals the energy needed to create an exciton and is given in the last row. The parameters taken from \cite[Table I]{Hawke2010} and given in $100\,\mathrm{meV}$.}
\begin{ruledtabular}
\begin{tabular}{l|cccc}
&A&T&G&C\\
\hline
LUMO (electrons)&-44&-49&-45&-43\\
HOMO (holes)&-83&-90&-80&-88\\
$\Delta E$ (exciton)& 39&41&35&45
\end{tabular}
\end{ruledtabular}
\end{table}

\begin{table}[H]
\caption{\label{tab:3} Tight-binding parameters for the direct interstrand transfer $h_e$ ($h_h$) of electrons (holes). The parameters taken from \cite[Table IV]{Hawke2010} and given in $\mathrm{meV}$.}
\begin{ruledtabular}
\begin{tabular}{cccc}
     A-T&-9 (-12)&G-C&16 (-12)
\end{tabular}
\end{ruledtabular}
\end{table}

\begin{table}[H]
\caption{\label{tab:2} Tight-binding parameters for the intrastrand transfer $t_e$ ($t_h$) of electrons (holes). The table is to be read in such a way that the parameter in row $X$ and column $Y$ corresponds to the transfer parameter from base $X$ to base $Y$ in 5'-3' direction (or the transfer parameter from base $Y$ to base $X$ in 3'-5' direction). The parameters are taken from \cite[Table V]{Hawke2010} and given in $\mathrm{meV}$.}
\begin{ruledtabular}
\begin{tabular}{l|cccc}
     &A&T&G&C\\
     \hline
     A&16 (-8)&7 (68)&1 (-5)&-3 (68)\\
     T&-7 (26)&-30 (-117)&-17 (28)&22 (-86)\\
     G&30 (-79)&-32 (73)&20 (-62)&43 (80)\\
     C&-12 (5)&63 (-107)&15 (-1)&-47 (-66)
    \end{tabular}
\end{ruledtabular}
\end{table}

\begin{table}[H]
\caption{\label{tab:4} Tight-binding parameters for the diagonal interstrand transfer $r_e^{+}$ ($r_h^{+}$) of electrons (holes) in 5'-5' direction. The table is to be read in such a way that the parameter in row $X$ and column $Y$ corresponds to the transfer parameter from base $X$ to base $Y$ in 5'-5' direction (or the transfer parameter from base $Y$ to base $X$ in 3'-3' direction). The parameters taken from \cite[Table VII]{Hawke2010} and given in $\mathrm{meV}$.}
\begin{ruledtabular}
\begin{tabular}{l|cccc}
    &A&T&G&C\\
    \hline
    A&6 (2)&2 (9)&3 (4)&-2 (5)\\
    T&2 (9)&2 (4)&3 (5)&-2 (2)\\
    G&3 (4)&3 (5)&-2 (3)&-3 (4)\\
    C&-2 (5)&-2 (2)&-3 (4)&2 (1)
\end{tabular}
\end{ruledtabular}
\end{table}

\begin{table}[H]
\caption{\label{tab:5} Tight-binding parameters for the diagonal interstrand transfer $r_e^{-}$ ($r_h^{-}$) of electrons (holes) in 3'-3' direction. The table is to be read in such a way that the parameter in row $X$ and column $Y$ corresponds to the transfer parameter from base $X$ to base $Y$ in 3'-3' direction (or the transfer parameter from base $Y$ to base $X$ in 5'-5' direction). The parameters taken from \cite[Table VI]{Hawke2010} and given in $\mathrm{meV}$.}
\begin{ruledtabular}
\begin{tabular}{l|cccc}
    &A&T&G&C\\
    \hline
    A&29 (48)&3 (-3)&-6 (-3)&-3 (-5)\\
    T&3 (-3)&0.2 (0.5)&2 (5)&-0.2 (0.5)\\
    G&-6 (-3)&2 (5)&-5 (-44)&-4 (4)\\
    C&-3 (-5)&-0.2 (0.5)&-4 (4)&0.3 (1)
\end{tabular}
\end{ruledtabular}
\end{table}

\subsection{Wire model}

\begin{table}[H]
\caption{\label{tab:6} HOMO and LUMO energies of paired DNA bases. The HOMO energies are taken as on-site energies $\epsilon_h$ of the holes and the LUMO energies as on-site energies $\epsilon_e$ of the electrons. The energy difference $\Delta E$ between HOMO and LUMO equals the energy needed to create an exciton and is given in the last row. The parameters taken from \cite[Table I]{Simserides2014} and given in $100\,\mathrm{meV}$.}
\begin{ruledtabular}
\begin{tabular}{l|cccc}
&A-T&G-C\\
\hline
LUMO (electrons)&-49&-45\\
HOMO (holes)&-83&-80\\
$\Delta E$ (exciton)& 34&35\\
\end{tabular}
\end{ruledtabular}
\end{table}

\begin{table}[H]
\caption{\label{tab:7} Tight-binding parameters for the intrastrand transfer $t_e$ ($t_h$) of electrons (holes). The table is to be read in such a way that the parameter in row $X$ and column $Y$ corresponds to the transfer parameter from base $X$ to base $Y$ in 5'-3' direction (or the transfer parameter from base $Y$ to base $X$ in 3'-5' direction). The parameters taken from \cite[Table II]{Simserides2014} and given in $\mathrm{meV}$.}
\begin{ruledtabular}
\begin{tabular}{l|cccc}
     &A&T&G&C\\
     \hline
     A& -29 (20) & 0.5 (-35) & 3 (30) & 32 (-10)\\
     T& 2 (-50) & -29 (20) & 17 (10) & -1 (110) \\
     G& -1 (110) & 32 (-10) & 20 (100) & -10 (-10)\\
     C& 17 (10) & 3 (30) & -8 (50) & 20 (100)
    \end{tabular}
\end{ruledtabular}
\end{table}


\section{Extra plots}

In this section, we provide additional plots and data that complement and extend the analyses presented in the main text. These supplementary figures aim to offer a more comprehensive understanding of the key findings discussed in the primary research manuscript.

\begin{figure*}
\centering
\includegraphics[width=1\linewidth]{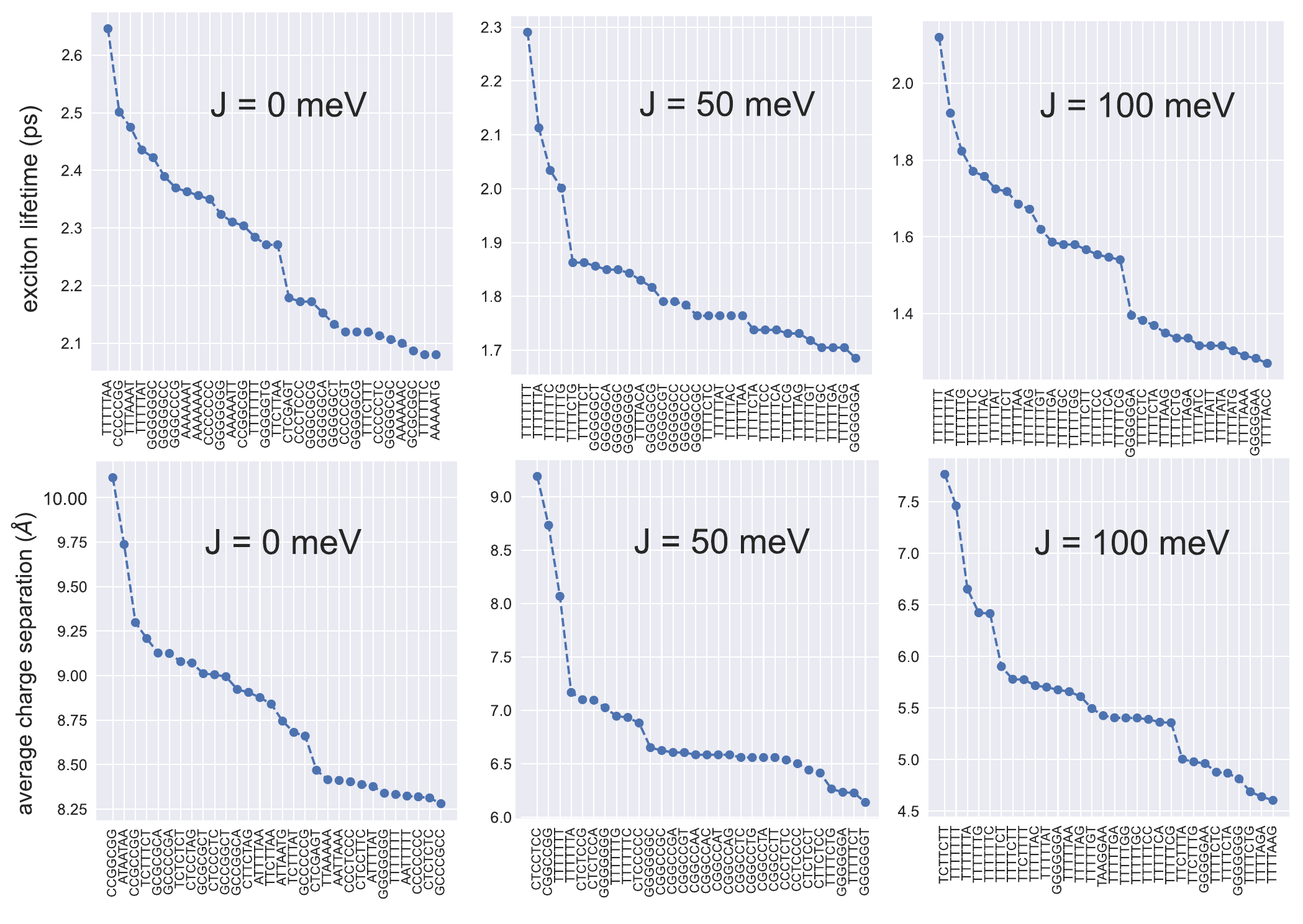}
\caption{\label{fig:S1} Top 30 DNA sequences showing the highest exciton lifetime (upper row) and average charge separation (lower row) for no interaction, intermediate interaction ($J=50\,\mathrm{meV}$) and strong interaction ($J=100\,\mathrm{meV}$).}
\end{figure*}

\begin{figure*}
\centering
\includegraphics[width=1\linewidth]{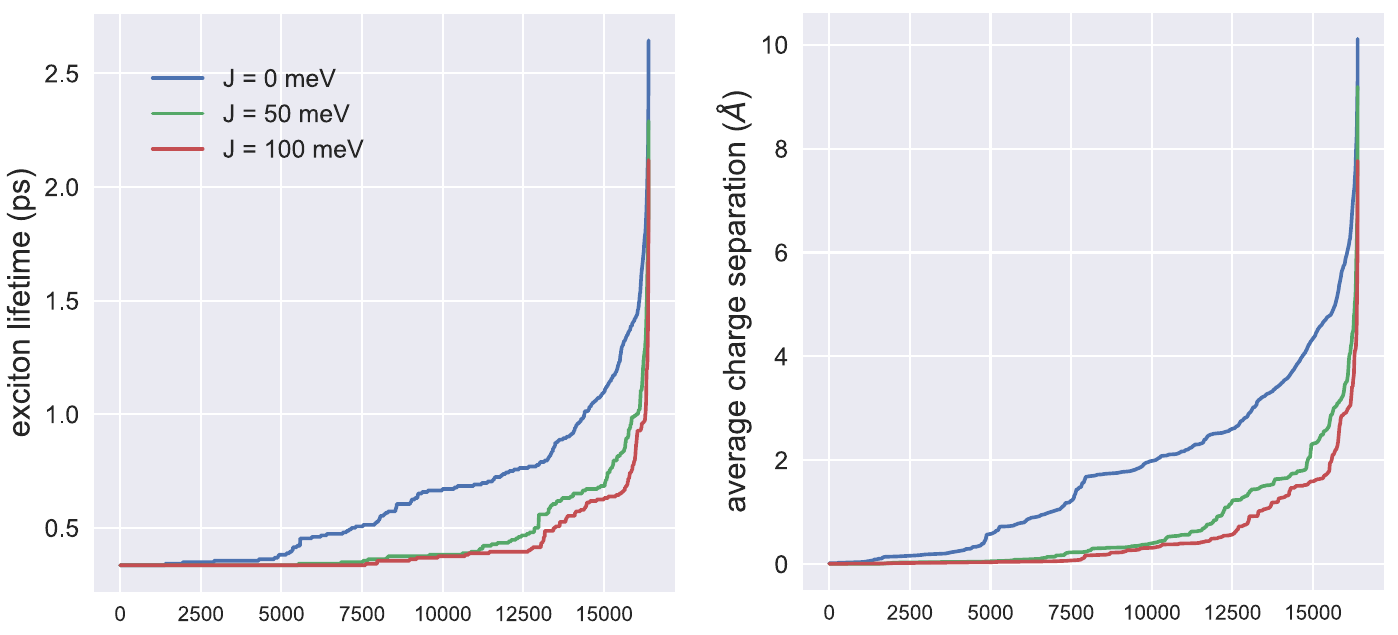}
\caption{\label{fig:S2} Results of all sequences consisting of seven base pairs ordered by exciton lifetime (left) and average charge separation (right) for no interaction, intermediate interaction ($J=50\,\mathrm{meV}$) and strong interaction ($J=100\,\mathrm{meV}$).}
\end{figure*}

\begin{figure*}
\centering
\includegraphics[width=0.9\linewidth]{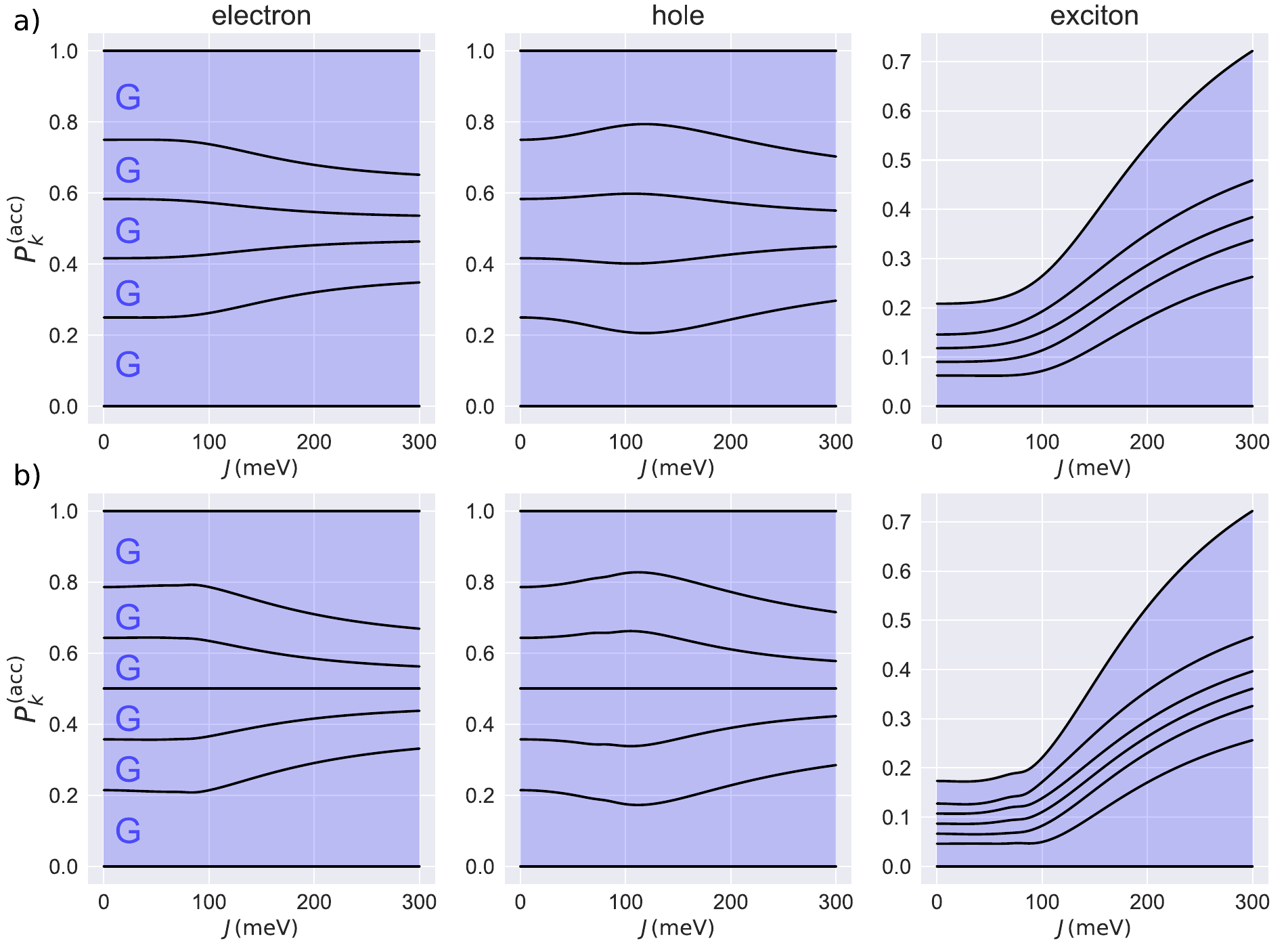}
\caption{\label{fig:S3} Accumulated average populations $P_k^{\text{acc}}=\sum_{i=0}^k\langle P_i\rangle$ of (a) GGGGG and (b) GGGGGG with initial excitation on the left base pair according to the wire model description. The hopping parameters are as follows: $t_e=20\,\mathrm{meV}$ for the electron and $t_h=100\,\mathrm{meV}$ for the hole.}
\end{figure*}

\begin{figure*}
\centering
\includegraphics[width=1.0\linewidth]{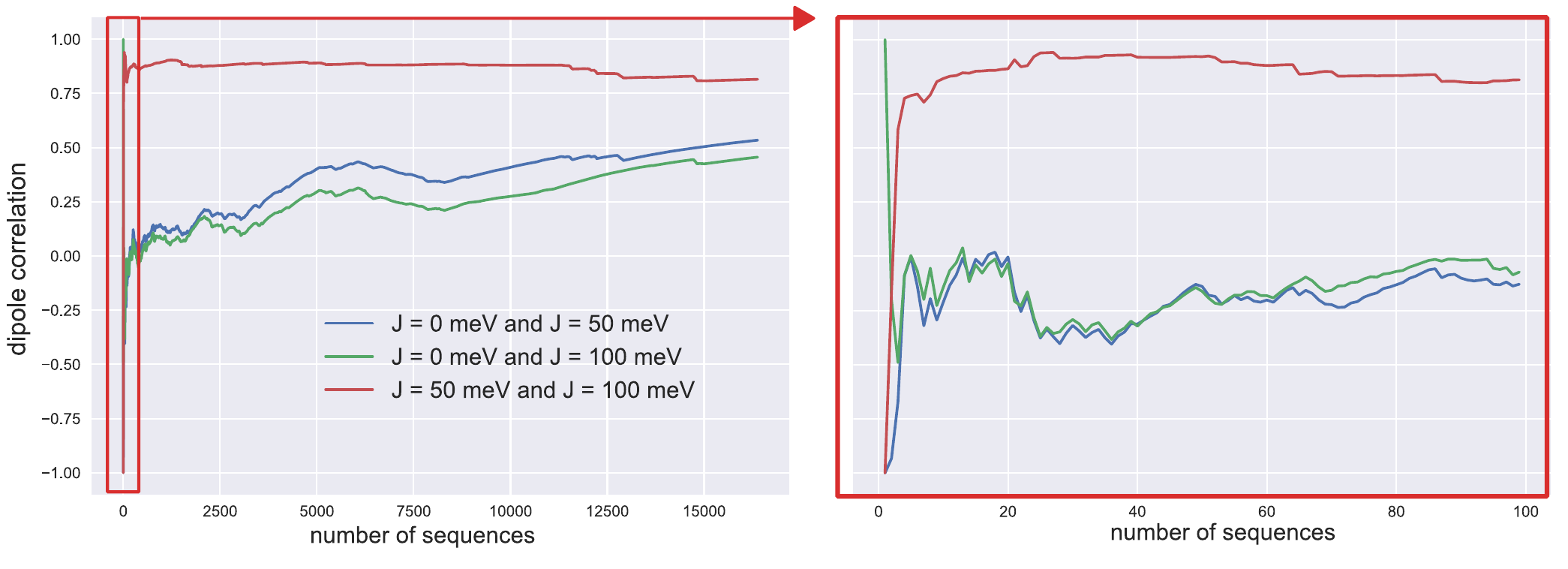}
\caption{\label{fig:S4} Correlation plots between lists of DNA sequence dipole moments, for different values of the e-h interaction strength (see legend). The right plot is a zoom into the first 100 sequences of the left plot. The correlation between different non-zero e-h interactions is always higher than the correlations with zero e-h interaction. }
\end{figure*}

In particular, we present the results of our analysis of the average charge separation between electrons and holes in the time interval between photoexcitation and relaxation of the DNA double strand. Fig.~S\ref{fig:S1} shows that, as discussed in the main text, the composition of the top 30 sequences changes with increasing interaction strength toward a higher amount of thymine (T). This is true for both the exciton lifetime and the average charge separation, although the sequences themselves are different in both analyses. In Fig.~S\ref{fig:S2} we show all DNA sequences of seven base pairs ordered by exciton lifetime (left) and average charge separation (right) for selected interaction strengths. We find that for the vast majority of DNA sequences (more than 10,000 sequences, especially strongly non-symmetric and non-uniform) the excitation decays on the same time scale as for an isolated single base and the charges show almost no separation, supporting the well-known result that DNA is photostable. Fig.~S\ref{fig:S3} shows the same kind of analysis on the mean-over-time population with varying interaction strength as in Fig.~5 of the main text, performed on uniform guanine sequences. Fig.~S\ref{fig:S4} shows the correlations for the dipole moment in the same way as in Fig.~7 of the main manuscript. 

\begin{figure*}
\centering
\includegraphics[width=0.8\linewidth]{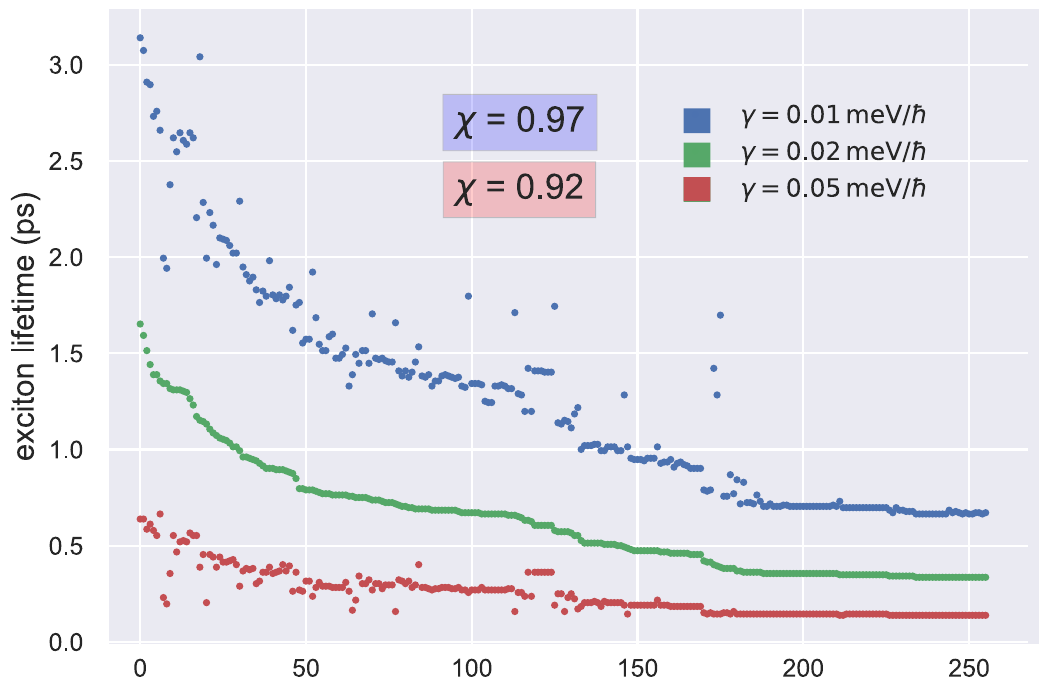}
\caption{\label{fig:S5}Excitonic lifetime plots for S$_4$ sequences with different relaxation rates for the release of energy into the environment. For different rates of excitonic escape, the different resulting sequences show strong correlations and very similar behavior, confirming the robustness of this phenomenon to this parameter. The correlation of $\chi = 0.97$ shown refers to the two sequences with $\gamma = 0.01 - 0.02$ and that of $\chi = 0.92$ refers to the two sequences with $\gamma = 0.05 - 0.02$.}
\end{figure*}